\documentclass[letterpaper]{article} 
\usepackage{aaai25}  
\usepackage{times}  
\usepackage{helvet}  
\usepackage{courier}  
\usepackage[hyphens]{url}  
\usepackage{graphicx} 
\usepackage{natbib}  
\usepackage{caption} 
\usepackage{algorithm}
\usepackage{algorithmic}

\usepackage{amssymb}
\usepackage{bm}
\usepackage{amsmath}
\usepackage{dsfont}
\usepackage{booktabs}
\usepackage{multirow}

\usepackage{newfloat}
\usepackage{listings}
\DeclareCaptionStyle{ruled}{labelfont=normalfont,labelsep=colon,strut=off} 
\floatstyle{ruled}
\newfloat{listing}{tb}{lst}{}
\floatname{listing}{Listing}
\title{Emotion-Agent: Unsupervised Deep Reinforcement Learning with Distribution-Prototype Reward for Continuous Emotional EEG Analysis}

\author{
        Zhihao Zhou\textsuperscript{\rm 1,2}, Qile Liu\textsuperscript{\rm 1,2}, Jiyuan Wang\textsuperscript{\rm 1,2}, Zhen Liang\textsuperscript{\rm 1,2,}\thanks{Corresponding author.} \\
}
\affiliations{
    \textsuperscript{\rm 1}School of Biomedical Engineering, Shenzhen University, Shenzhen, 518060, Guangdong, China \\
    \textsuperscript{\rm 2}Guangdong Provincial Key Laboratory of Biomedical Measurements and Ultrasound Imaging, Shenzhen, China \\
    \{2310247057, liuqile2022, 2310247016\}@email.szu.edu.cn, janezliang@szu.edu.cn
}

\begin{document}

\maketitle

\begin{abstract}
Continuous electroencephalography (EEG) signals are widely used in affective brain-computer interface (aBCI) applications. However, not all continuously collected EEG signals are relevant or meaningful to the task at hand (e.g., wondering thoughts). On the other hand, manually labeling the relevant parts is nearly impossible due to varying engagement patterns across different tasks and individuals. Therefore, effectively and efficiently identifying the important parts from continuous EEG recordings is crucial for downstream BCI tasks, as it directly impacts the accuracy and reliability of the results. In this paper, we propose a novel unsupervised deep reinforcement learning framework, called Emotion-Agent, to automatically identify relevant and informative emotional moments from continuous EEG signals. Specifically, Emotion-Agent involves unsupervised deep reinforcement learning combined with a heuristic algorithm. We first use the heuristic algorithm to perform an initial global search and form prototype representations of the EEG signals, which facilitates the efficient exploration of the signal space and identify potential regions of interest. Then, we design distribution-prototype reward functions to estimate the interactions between samples and prototypes, ensuring that the identified parts are both relevant and representative of the underlying emotional states. Emotion-Agent is trained using Proximal Policy Optimization (PPO) to achieve stable and efficient convergence. Our experiments compare the performance with and without Emotion-Agent. The results demonstrate that selecting relevant and informative emotional parts before inputting them into downstream tasks enhances the accuracy and reliability of aBCI applications. 
\end{abstract}

%

\section{Introduction}
\label{sec:Introduction}
{Human emotion is a continuous dynamic process, characterized by complex interactions between both internal and external components of the human body \cite{cowen2017self,horikawa2020neural}. How to identify task-related emotional segments from continuous EEG signals presents a significant challenge. Electroencephalography (EEG) provides a direct, objective, and scientifically grounded method for assessing emotional states, making it a valuable tool in emotion recognition research \cite{song2018eeg}. In recent years, the potential of EEG-based emotion recognition has garnered increasing attention from researchers across diverse disciplines \cite{li2021multi,gong2023astdf,liu2024joint}.}

{One significant limitation of existing research is the reliance on a static labeling approach, where a single, fixed label is assigned to an entire EEG segment. This method fails to capture the dynamic nature of human emotions during EEG-evoked experiments, as emotional states are inherently fluid, constantly shifting in response to both internal cognitive processes and external stimuli \cite{huang2014novel,liu2017real}. Moreover, continuous EEG recordings often include states that are irrelevant to the specific task being studied. These irrelevant states can introduce noise and confounding factors, undermining the accuracy and reliability of emotion recognition models. Current methods face challenges in isolating and identifying the task-related moments within the EEG data that are most relevant to the study. When task-irrelevant EEG segments are included in the training data, they introduce extraneous information that can degrade the model's performance. As a result, the model may mistakenly associate these irrelevant patterns with emotional states, leading to reduced accuracy in emotion recognition by diverting attention from the true task-related emotional dynamics \cite{li2019domain,zheng2015investigating,zheng2016multichannel}. On the other hand, requiring real-time annotation of task-related segments during an experiment is impractical. This is especially true when considering that wandering thoughts or irrelevant mental states are often indistinguishable even to the subject themselves. Thus, developing an artificial intelligence (AI) empowered method that can dynamically adapt to the fluid nature of human emotions and accurately isolate task-relevant EEG segments is essential for improving the precision and effectiveness of emotion recognition models.}

{Deep reinforcement learning, with its adaptability and flexibility in uncertain environments, offers a promising solution to this challenge \cite{vinyals2019grandmaster,kalashnikov2018scalable}. By leveraging a reward-based mechanism, it reduces the dependence on labels and enables unsupervised autonomous exploration of task-relevant information. For example, Zhou \cite{zhou2018deep} proposed a Diversity-Representativeness Reward to guide Agent in generating more diverse and representative video summaries. Similarly, AC-SUM-GAN \cite{apostolidis2020ac} used an Actor-Critic framework to exploit the reconstruction error of the discriminator as a reward function, with the Critic guiding the Actor through gradient feedback to learn strategies for extracting key video segments. In the field of EEG emotion computing, TAS-Net \cite{zhang2023unsupervised} proposed the use of deep reinforcement learning to detect the most informative key emotional segments from EEG signals in an unsupervised manner. However, these methods often fail to incorporate information about the overall distribution of EEG signal features and struggle to capture the long-term similarities in human emotions. This can lead to gaps in understanding the continuity and subtle shifts in emotional states, potentially affecting the accuracy and effectiveness of emotion recognition models.}

{To address the limitations of existing research, we formulate the extraction of key EEG segments as a sequential decision-making process and introduce a novel Emotion-Agent designed to automatically identify relevant and informative emotional moments from continuous EEG signals. Emotion-Agent integrates reinforcement learning with heuristic search algorithms to enhance the RL agent's exploration process during training. By utilizing the efficient search capabilities of heuristic algorithms, the agent can minimize exploration of low-value trajectories, making the process more targeted and purposeful. Consequently, the model achieves more efficient convergence, even when accounting for the inherent costs of exploration. The proposed Emotion-Agent is capable of effectively capturing the most significant emotional segments without the need for predefined labels. The main contributions of this paper are summarized as follows.}
\begin{itemize}
    \item We propose a novel Emotion-Agent, which integrates deep reinforcement learning with heuristic algorithms to optimize the extraction of key emotional segments.
    \item The reward function, named Distribution-Prototype, is designed with a focus on distribution, considering both local and global sample distributions during the reward learning process.
    \item Extensive experimental results demonstrate that selecting relevant and informative emotional segments enhances the accuracy and reliability of emotion analysis.
\end{itemize}
\section{Related Work}
\subsection{Reinforcement Learning}
{Reinforcement Learning (RL) is a powerful machine learning paradigm where an intelligent agent learns an optimal decision policy by interacting with its environment \cite{zoph2017neuralarchitecturesearchreinforcement}. Unlike other machine learning methods, RL emphasizes learning through trial and error, with the agent taking actions to maximize cumulative rewards over time. This approach has gained significant traction across various domains due to its ability to handle complex, dynamic environments where the agent’s decisions continuously adapt based on new information \cite{he2016deepreinforcementlearningnatural,yarats2021image}.
\subsection{Reinforcement Learning with Heuristics}
Heuristic-Guided Reinforcement Learning (HuRL) was introduced \cite{cheng2021heuristic}, aiming to accelerate traditional RL algorithms by incorporating heuristics derived from domain knowledge or offline data. These heuristics guide the RL agent, enabling more informed decisions and speeding up the learning process. HuRL is particularly valuable in environments where the state space is vast, making unguided exploration computationally expensive and time-consuming. Another significant advancement is the introduction of large-state reinforcement learning for hyper-heuristics \cite{kletzander2023large}. This approach leverages solution change trajectories from an extensive feature set, integrating them into the RL framework. By incorporating local search principles and introducing a probability distribution within the $\epsilon$-greedy strategy, this method increases the likelihood of sampling high-quality sequences of low-level heuristics. It significantly enhances the efficiency of RL in solving complex optimization problems with exceptionally large state spaces. A novel solution for continuous trajectory generation in urban road networks was also proposed \cite{jiang2023continuous}, combining a two-stage Generative Adversarial Network (GAN) with A* heuristic search algorithms. This design features discriminators for sequential reward and movement yaw reward, guiding the agent in generating more accurate and efficient trajectories. Building on the foundations of RL, personalized reinforcement learning was introduced \cite{ivanov2024personalized}. Inspired by the classical K-means clustering principle, this approach incorporates the concept of a budget of policies within robust Markov Decision Processes (r-MDPs). The framework enables the RL agent to interact with users through representative policies, efficiently adapting to individual user preferences. An earlier application of RL in the field of education is demonstrated with AgentX \cite{martin2004agentx}. This intelligent agent was developed to enhance the effectiveness of Intelligent Tutoring Systems (ITS). By clustering personalized group information about students, the RL-based AgentX tailors the learning experience for each group.

The advancements in reinforcement learning across various domains: from accelerating traditional RL algorithms with heuristics to personalizing user interactions and improving intelligent tutoring systems—demonstrate the versatility and potential of this machine learning paradigm. The continuous evolution of RL, as seen in large-state optimization and urban trajectory generation, underscores its capacity to tackle increasingly complex challenges. As RL continues to integrate with other AI techniques, such as GANs and heuristic search algorithms, it is positioned to drive significant innovations across a wide array of fields, shaping the future of intelligent systems and autonomous decision-making \cite{warnell2018deep,ren2023leveraging,vecerik2017leveraging}.

\section{{Methodology}}{
\subsection{Markov Decision Process}
\label{subsec:Problem Definition}{

{We model the detection of the most emotionally relevant segments from sequential EEG signals as a sequential decision-making process, formulated as a Markov Decision Process (MDP). An MDP is defined as a tuple $\mathcal{M}=\mathcal{<}\mathcal{S}, \mathcal{A}, \mathcal{P}, \mathcal{R}\mathcal{>}$, where $\mathcal{S}$ represents the state space, and $\mathcal{A}$ denotes the action space, with $\mathcal{A}=\{0,1\}$ corresponding to the possible actions the agent can take. The transition probability function $\mathcal{S} \times \mathcal{A} \times \mathcal{S} \rightarrow [0,1]$ describes the likelihood of transitioning from one state to another given a specific action. The reward function $\mathcal{R}: \mathcal{S}\times\mathcal{A}\times\mathcal{S}\rightarrow\mathcal{R}$ assigns a numerical reward based on the state-action-state transition, providing feedback on the agent's decisions.}

{At each timestep $t$, the RL agent, upon executing an action $a_t$ in state $s_t$, transitions to a new state $s_{t+1}$ and receives a corresponding reward $r_t$. We define $\mathcal{M}$ tuple as transitions. In an episode when interacting with the environment, we collect multiple trajectories consisting of multiple transitions. Through repeated interactions with the environment, the RL algorithm aims to learn an optimal policy $\pi$ that maximizes the cumulative reward over time. This optimal policy enables the agent to make decisions that consistently lead to the identification of key emotional segments in the EEG signals. }
}

\begin{figure*}[h]
    \begin{center}
        \includegraphics[width=1.0\textwidth]{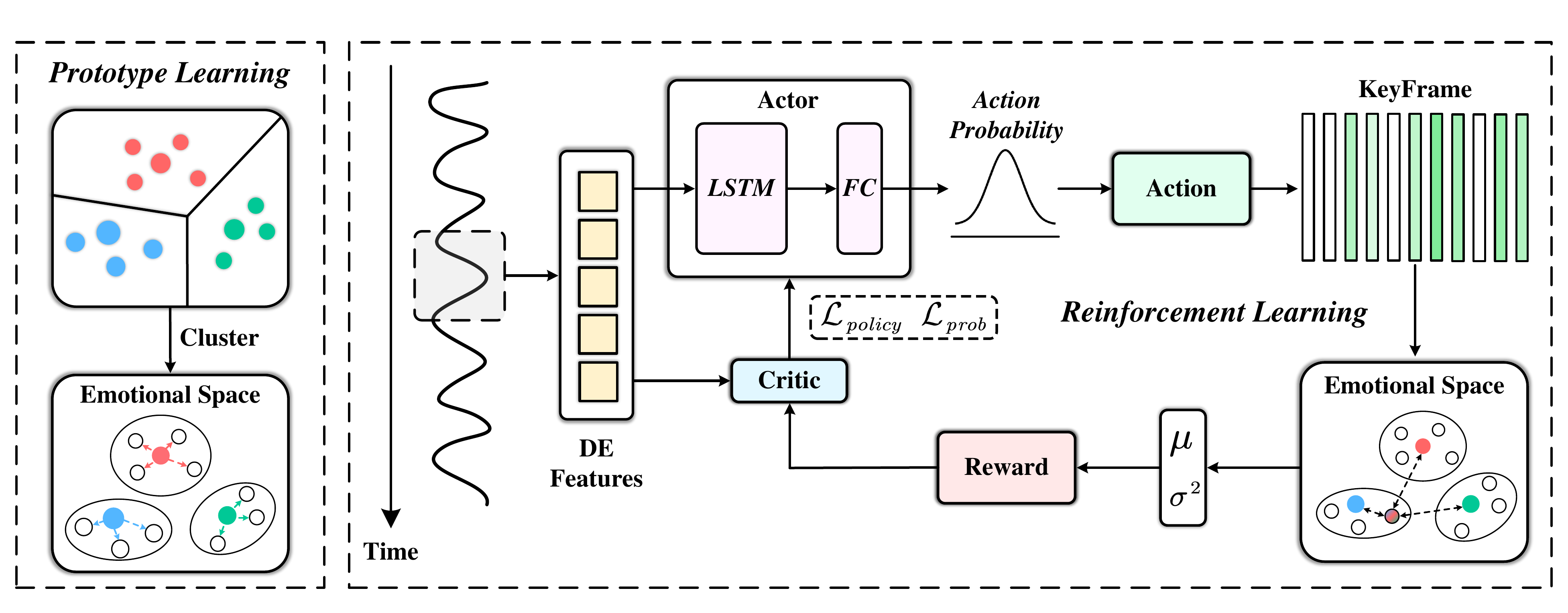}
    \end{center}
    \caption{The framework of the proposed Emotion-Agent.The Emotion Agent is divided into two stage in total: (1) Heuristic global search for prototype learning stage, and (2) Reinforcement learning stage based on Distribution-Prototype reward}
    \label{fig:Visualization}
\end{figure*}

\subsection{Prototype Learning}
\label{subsec:Model Architecture}{
{In the process of extracting key segments from task-related EEG signals, it is crucial to consider not only the intrinsic data distribution characteristics of each segment but also the broader context provided by the global emotion space, which encapsulates the distribution of various emotion categories. To achieve a more effective representation of EEG emotion distribution, we introduce the concept of prototype learning\cite{zhou2023pr}. This approach allows us to model each emotion category as a prototype, thereby capturing the globally distributed emotional information with better representation. Prototype learning enables us to integrate global emotion information into the reward structure, ensuring that the reinforcement learning process is informed by a comprehensive understanding of the emotional landscape represented in the EEG data.}

{Specifically, we employ the K-Means clustering algorithm as a heuristic method. This step enables the model to obtain a global perspective on the distribution of emotional information across all subjects' EEG data. By inputting the differential entropy (DE) features of EEG signals from all subjects into the K-Means algorithm, we derive the set of emotion prototypes $\{C^{i}\}_{i=1}^{N}$, where each $C^i$ represents a cluster center corresponding to a emotion category, $N$ represents the number categories. By clustering the data, we identify the optimal emotion prototypes that serve as representative points in the emotion space. These prototypes are then used to inform the design of the reward function, ensuring that it reflects the global distribution of emotions captured in the EEG signals. The prototype feature vector for a given emotion category $c$ can be calculated by averaging all the sample features that belong to this category. Mathematically, the prototype feature vector $\mu_i$ is given by:}
\begin{equation}
\label{eq:cluster center}
\mu_{c}=\frac{1}{|C^i|}\sum_{x_{i}\in C^{i}}f\left(x_{i}\right),
\end{equation}
{where $C^{i} = \{(x_{i}, y_{i} = c)\}_{i=1}^{N}$ represents the set of samples belonging to the emotion category $c$, and $|C^{i}|$ is the number of samples in this category. The centroid $\mu_{c}$ serves as the average feature vector for the emotion category $c$.}

{The K-Means algorithm iteratively reclassifies data points and updates cluster centers to minimize the sum of squared errors within the clusters. The objective function of the algorithm is defined as:}
\begin{equation}
\arg\operatorname*{min}_{C}\sum_{i=1}^{N}\sum_{x \in C^{i}}\|x-\mu_{c}\|_{2}.
\end{equation}
}
{To quantify the variance within each cluster, we use the mean of the sum of squared intra-cluster errors, which reflects the distribution of EEG features within the cluster. The variance of the intra-cluster distribution is indicative of the individual variability of each emotion. The intra-cluster variance $\sigma^{2}_{i}$ is calculated using the following formula:}
\begin{equation}
\sigma^{2}_{i} = \frac{1}{|C^i|}\sum_{x_i \in C^i}\|x_{i}-\mu_{{c}}\|_2,
\end{equation}
{where $|C^i|$ is the number of data points in the cluster, $x_{i}$ represents an individual data point, and $\mu_{{c}}$ is the corresponding cluster center.}

{To better represent the distribution of emotions in the EEG data, the prototype learning process integrates the cluster centers as the mean of the data distribution and uses the mean of the sum of squared errors within clusters to describe the variance. This combination of prototype learning and K-Means clustering provides a robust foundation for the reinforcement learning process, enabling the model to effectively navigate and interpret the complex emotional information present in EEG data.}

\subsection{Distribution-Prototype Reward}
{We obtain a global distribution of emotional information in an unsupervised manner through prototype learning, and we incorporate this global information into the reward function for our reinforcement learning model. We believe that the clustering centers obtained through heuristic search represent the prototypes of each affective category. These prototypes effectively capture the distribution of sample features across the entire affective category, with other EEG features belonging to the same category clustering around these prototypes.}

{From a probabilistic perspective, an emotion prototype can be understood as the mean of the emotion sample features, while the variance in this distribution arises from the inherent variability of human emotions and the non-stationarity of EEG signals. We use the mean and variance of each emotion cluster to reflect the individual variability within the distribution of EEG features for each emotion category. The goal of the Actor in our proposed method is to maximize the expected reward over time by selecting key segments that are more closely aligned with the target emotion. To achieve this, we propose two reward functions based on the distributional information derived from prototype learning: center reward and inter-intra reward.}

\textbf{Center reward}
{This reward function measures the intensity of the emotional information contained in the features of the current EEG sample. We evaluate this intensity by calculating the Euclidean distance between the sample point and the cluster center:}
\begin{equation}
    Reward_{center}=\frac{1}{1+dist} ,
\end{equation}
Where $dist$ calculate the distance from the EEG sample feature to the clustering centre of the category it belongs to.
\begin{equation}
    \large dist=||x_{c^i}-\mu_{c^i}||_2 .
\end{equation}

{We use this distance measure between sample points and cluster centers in the emotion space for two purposes. First, we consider that the proximity of sample feature points to cluster centers reflects the intensity of the emotion they represent. Second, it helps to mitigate the effect of interfering information from outliers in the EEG signal caused by non-stationarity.}

\textbf{Inter-Intra reward}
{This reward function represents the confidence level that the current EEG sample feature belongs to the specific emotion category. We employ inverse variance weighting to calculate the distances between EEG sample features and the cluster centers of other categories. This method provides a weighted mean with the smallest variance \cite{lin2021combining}, which we use to estimates the confidence $Inter$ that the sample point is far from the centroid of the other category.
}
\begin{equation}
    Reward_{inter-intra} = exp(-\dfrac{Intra}{Inter}) ,
\end{equation}
\begin{equation}
\label{intra}
    Intra = ||x_{c^i}-\mu_{c^i}||_2,
\end{equation}
\begin{equation}
    Inter=\frac{\sum_{{x_i\notin C^i}}\frac{\left\|x_i-x_{{c^i}}\right\|_2}{\frac{1}{\sigma^2}}}{\sum_{{x_i\notin C^i}}\frac1{\sigma_i^2}} ,
\end{equation}

{where $Intra$ estimates the confidence which the sample point belongs to the target category.}

\subsection{Optimization Process}
{We use PPO \cite{schulman2017proximal} to train our model, as it searches for emotionally relevant EEG key segments at the trial level. Throughout this process, the model learns to identify emotionally prototypical policies in a trial-and-error manner within a discrete action space. To stabilize the training process, we employ PPO-Clip, which restricts the ratio between old and new policies, reducing oscillations and accelerating convergence. PPO is a policy-based Actor-Critic method. To improve sample efficiency in the On-Policy training process, importance sampling is introduced, allowing the model to reuse trajectories multiple times:}
\begin{equation}
\label{equation: ratio}
    r_t(\theta)=\frac{\pi_\theta(a_t|s_t)}{\pi_{\theta_{\mathrm{old}}}(a_t|s_t)} ,
\end{equation}
{where $r_t(\theta)$ represents the probability ratio between the current and previous policies.}

{To better estimate cumulative returns, we use Generalized Advantage Estimation (GAE), which provides a more accurate advantage function. The specific expression of GAE is as follows:}
\begin{equation}
\label{equation: GAE}
    \hat{A_t}^{ GAE(\gamma,\lambda)}=\sum_{l=0}^\infty(\gamma\lambda)^l\delta_{t+1}^V=\delta_t^V+\gamma\lambda\hat{A}_{t+1}^{GAE(\gamma,\lambda)} ,
\end{equation}
{where $\gamma$ is the discount factor and $\lambda$ is the GAE hyperparameter that controls the trade-off between bias and variance. GAE uses a weighted average of multiple value estimates, and to quickly estimate the advantage at each time step, a recursive calculation is performed, estimating time $t$ from time $t+1$. To further stabilize the training process, we apply PPO-Clip, which limits the changes between the old and new policies. The objective function with clipping can be expressed as:}
\begin{equation}
\label{equation: object}
\mathcal{L}_{policy}(\theta)=\hat{E}_t\left[\min\left(r_t(\theta)\hat{A}_t,\mathrm{clip}(r_t(\theta),1-\epsilon,1+\epsilon)\hat{A}_t\right)\right] ,
\end{equation}
{where $\epsilon$ control the update range of the action probability at each iteration by setting upper and lower thresholds on the ratio of the new and old strategies.
}

\subsection{Regularization}
{To prevent the Actor from selecting too many keyframes during an episode, we introduce a regularization term that constrains the action probability of the learned policy function. This can be expressed as:}
\begin{equation}
    L_{prob}=||\frac1T\sum_{t=1}^Tp_t-\delta||^2,
\end{equation}
{where $\delta$ is a scalar representing the desired proportion of key emotional segments selected. We then use the Adam optimizer to update the parameters $\theta$ of the policy function, calculated as:}
\begin{equation}
    \theta=\theta-\varphi\nabla_{\theta}\left(-\mathcal{L}_{policy}+\beta\mathcal{L}_{prob}\right) ,
\end{equation}
{where $\varphi$ is the learning rate, and $\beta$ is a regularization coefficient.}

{The Critic network is responsible for estimating $V(S_t)$ during the decision process, which serves as a prediction of the actual discounted cumulative reward throughout the process. This estimation guides the Actor network towards converging on the optimal policy. The error function for the Critic is the mean square error (MSE) between the estimated value and the actual discounted cumulative reward, and is expressed as:}
\begin{equation}
    L(\theta)=\mathbb{E}\left[\left(V(s_t)-R_t\right)^2\right] ,
\end{equation}
{where $R_t$ is the discounted cumulative reward at time $t$. The Critic's predicted value is then used to calculate the MSE loss relative to the actual discounted reward.}

{During the training process, the model is divided into two parts: Actor and Critic. The Actor continually interacts with the emotion space during decision-making, iteratively searching for an optimal strategy based on the reward mechanism provided by environmental feedback. The Critic guides this exploration process by estimating the cumulative reward for each state, thereby assisting the Actor in finding the optimal strategy.}

\begin{algorithm}[h]

    \caption{The pre-training process of Emotion-Agent.}
    \label{alg:Pre-training Process of SCMM}
    \textbf{Input}: DE feature sequence $\{s_{t}\}^{T}_{t=1}$ from trainging set\\
    \textbf{Output}: Parameters $\theta$ of the Emotion-Agent.
    
    \begin{algorithmic}[1]
    \STATE Input All Subjects DE feature data into K-Mean, calculate $\mu_i$ according to Eq.(2), calculate $\sigma^2$ according to Eq.(4);\\
    \STATE Initial actor parameters $\theta_0$, initial critic parameters $\phi_0$; \\
        \FOR{$ i \rightarrow 1,2,...,\boldsymbol{\xi}$ }
            \STATE Input a DE feature sequence $\{s_t\}^T_{t=1}$ from training set\\
            
            \STATE Collect set of trajectories $\mathcal{D}_k=\{\tau_i\}$ by policy $\pi_k$ of actor;
            \STATE Compute generalized advantage estimates $\hat{A}_{t}$, based on the current value function $V_{\phi k}$ ;
            \STATE Update the policy of actor by maximizing the PPO-Clip objective function Eq.(11);
            \STATE Update the value function of critic by regression on Mean-Squared Error;
        \ENDFOR \\
    \STATE Save the parameters $\theta$ of the Emotion-Agent.
    \end{algorithmic}
\end{algorithm}

\section{Experiments}
\label{sec:Experiments}{
\subsection{{Datasets and Implementation Details}}
\label{subsec:Datasets}{
{Extensive validation experiments are conducted two publicly available datasets, including SEED \cite{zheng2015investigating} and DEAP \cite{koelstra2011deap}). We use DE features as inputs for the model. The details of the dataset and preprocessing will be introduced in the appendix.}
}

\begin{table*}[h]
    \centering
    \begin{tabular}{@{}c@{\hskip 0.2cm}|c@{\hskip 0.2cm}c@{\hskip 0.2cm}c@{}}
        \toprule
         Methods & Classification Task & $P_{acc}$  \\
        \midrule
        \textbf{\textit{Supervised}}& \textbf{\textit{Subject-Dependent}} & {\textbf{\textit{Video-Level LOOCV}}}\\
        \midrule
            \hspace{0.2cm}GSCCA \cite{zheng2016multichannel}  & Three-Class & 82.96 \\
            \hspace{0.2cm}DGCNN \cite{song2018eeg}  & Three-Class & 90.40  \\
            \hspace{0.2cm}RGNN \cite{zhong2020eeg} & Three-Class & 94.24  \\
        \midrule\textbf{\textit{Supervised with Transfer Learning}}& \textbf{\textit{Subject-Independent}} & {\textbf{\textit{Subject-Level LOOCV}}}\\
        \midrule
            \hspace{0.2cm}BiDANN \cite{li2018novel} & Three-Class & 83.28  \\
            \hspace{0.2cm}JDA \cite{li2019domain} & Three-Class & 88.28 \\
            \hspace{0.2cm}PR-PL \cite{zhou2023pr} & Three-Class & 93.06  \\
        \midrule\textbf{\textit{Supervised without Transfer Learning}}& \textbf{\textit{Subject-Independent}} & \multicolumn{2}{c}{\textbf{\textit{Subject-Level LOOCV}}}\\
        \midrule
        \hspace{0.2cm}JDA \cite{li2019domain} (source domain only) & Three-Class & 58.23 \\
        \midrule\textbf{\textit{Unsupervised}}& \textbf{\textit{Subject-Independent}} & \multicolumn{2}{c}{\textbf{\textit{Subject-Level LOOCV}}}\\
        \midrule
                \hspace{0.2cm}EEGFuseNet \cite{liang2021eegfusenet} & Three-Class & 42.04  \\
        \hspace{0.2cm}TAS-Net \cite{zhang2023unsupervised} & Three-Class & 52.99  \\
        \hspace{0.2cm}\textbf{Emotion-Agent (Ours)}  & Three-Class & \textbf{62.31}   \\
        \bottomrule
        \end{tabular}
\label{tab:Model performance on the SEED }
\caption{{Model performance (\%) for cross-subject emotion recognition on the SEED dataset.}}
\end{table*}

The network of actor consists of one layer of LSTM where the number of hidden layer nodes is 128 and two fully connected layers where the hidden nodes are from $256\rightarrow128, 128\rightarrow2$. The network setup of Critic is one layer of LSTM 128 and two fully connected layers where the hidden layer nodes in the fully connected layers are $256\rightarrow128$, $128\rightarrow1$. actor is optimised by the The optimisation is done by Adam's optimiser and the learning rate is set to 1e-4 and Critic's learning rate is done by Adam's optimiser and the learning rate is set to 1e-3. In addition the PPO algorithm has the gamma set to 0.98, the lmbda set to 0.95 and the eps set to 0.2 for the algorithms Eq. (\ref{equation: ratio}), Eq. (\ref{equation: GAE}), Eq. (\ref{equation: object}). All experiments are conducted using PyTorch 1.13.1 on an NVIDIA GeForce RTX 3090 GPU. More implementation details and parameter analysis are provided in the appendix.}

\subsection{{Evaluation Settings and Metrics}}

We use two experimental protocols to evaluate our approach. (1) Cross-Subject: \textbf{Subject-Independent}, \textbf{Subject-Level LOOCV}. We use subject-leave-one-out cross-validation to test the performance of our proposed model over cross-subjects. (2) Within-Subject: \textbf{Subject-Independent}, \textbf{Video-Level LOOCV}.Based on the above experimental scheme, we extract the most emotion-related segments on the proposed model, which are then used for subsequent model method analysis.

We conducted relevant experiments using the proposed model Emotion-Agent on SEED,  DEAP datasets, Emotion-Agent extracted the key segments related to emotions in an unsupervised manner through the guidance of reward function, we used the extracted key segments for downstream task modelling, the experiments compared with and without Emotion-Agent's Accuracy, F1-Scores two evaluation metrics.




\subsection{{Experimental Results}}
\label{subsec:Experimental Results on the SEED and DEAP datasets}
We compare the proposed Emotion-Agent with the current state-of-the-art methods. The comparison results for the three-classified emotion recognition (positive, neutral, negative) task on SEED are given in Table 1, where the methodology and the experimental protocol used are clearly stated. Overall, a supervised learning based approach yields better emotion recognition performance compared to an unsupervised learning based approach due to the use of label information for modelling, but such a trained model introduces label noise, and the model that completes the training actually learns that it is not really relevant to the emotion. In comparison to the unsupervised approach, the classification accuracy $P_{acc}$ with KNN reached 62.31\% after our model extracted the key segments, which is an improvement of 9.32\% compared to the results of TAS-Net for emotion recognition with the same KNN classifier. The experimental results show that through a well-designed reward function, our proposed method Emotion-Agent is better able to extract more relevant and richer EEG emotion segments on the SEED dataset, and from the experimental results we further argue this result.

\begin{table*}[h]
    \centering
    \begin{tabular}{@{}c@{\hskip 0.2cm}|c@{\hskip 0.2cm}c@{\hskip 0.2cm}c@{\hskip 0.2cm}c@{}}
        \toprule 
        \multirow{2}{*}{Methods}  & \multicolumn{2}{c}{Arousal}  & \multicolumn{2}{c}{Valence}\\
        & $P_{acc}$ & $P_f$ & $P_{acc}$ & $P_f$ \\
        \midrule
            \hspace{0.2cm} \textbf{\textit{Supervised}} & \multicolumn{2}{c}{\textbf{\textit{Subject-Dependent}}}& \multicolumn{2}{c}{\textbf{\textit{Video-Level LOOCV}}} \\
        \midrule
            \hspace{0.2cm}EMD \cite{zhuang2017emotion} & 71.99 & - & 69.10 & - \\
            
        \midrule
            \hspace{0.2cm} \textbf{\textit{Supervised}} & \multicolumn{2}{c}{\textbf{\textit{Subject-Independent}}}& \multicolumn{2}{c}{\textbf{\textit{Subject-Level LOOCV}}} \\
        \midrule
            \hspace{0.2cm}DGCNN\cite{song2018eeg}  & 61.10 & - & 59.29 & - \\
            \hspace{0.2cm}ATDD-LSTM\cite{liang2019unsupervised}  & 72.97 & - & 69.06 & - \\
        
        \midrule
            \hspace{0.2cm} \textbf{\textit{Unsupervised}} & \multicolumn{2}{c}{\textbf{\textit{Subject-Independent}}}& \multicolumn{2}{c}{\textbf{\textit{Subject-Level LOOCV}}} \\
        \midrule
            \hspace{0.2cm}EEG-FuseNet \cite{liang2021eegfusenet}  & 58.55 & 72.00 & 56.44 & 70.83 \\
            \hspace{0.2cm}TAS-Net \cite{zhang2023unsupervised}  & 60.51 & 72.64 & 57.84 & 71.80 \\
            \hspace{0.2cm}\textbf{Emotion-Agent(Ours)} & \textbf{75.06} & \textbf{73.88} & \textbf{57.91} & \textbf{55.43} \\
        \bottomrule
        \end{tabular}
\label{tab:Model performance on the DEAP }
\caption{{Model performance (\%) for cross-subject emotion recognition on the DEAP dataset.}}
\end{table*}

On the other hand, we dichotomised both Arousal and Valence for subjects emotional states on the DEAP dataset. Table 2 gives the experimental results of classifying both labels on the DEAP dataset and comparing them with other methods. Our proposed model performs the same on the task of classifying the emotional intensity of Arousal, and $P_{acc}$ is 14.55\% higher than TAS-Net, and comparing some supervised learning methods on Subjcet-Independent, Subject-Level LOOCV is 2.09\% higher than the current SOTA ATDD-LSTM. In addition our proposed method outperforms TAS-Net by 0.07\% in the metric $P_{acc}$ on the Valence emotional potency binary classification task. The results on this dataset show that the proposed method is able to extract the EEG segments of subjects in emotionally strong states in an unsupervised manner very well.

Overall, our proposed method achieves better results on both SEED and DEAP datasets. Moreover, on the SEED dataset the Emotion-Agent extracts the key EEG segments that are more relevant to emotions in an unsupervised manner thus improving the performance of the results on the subsequent emotion recognition task. In addition, on the DEAP dataset, for the task of emotion classification of Arousal labels, our method is based on our designed reward function, which measures the emotion intensity of EEG sample features and the emotion category to which they belong, and measures the interaction between EEG sample features and emotion prototypes well, and is able to accurately extract EEG emotion segments of people in emotionally intense moments in an unsupervised manner.
}

\section{Discussion}
\label{subsec:Discussion}{
To further validate our proposed Emotion-Agent model in terms of accuracy and reliability improvement in sentiment analysis, we conducted additional experiments on the SEED dataset. The experiments compared with and without, with i.e., using our proposed method Emotion-Agent extracts the key segments related to emotions and then inputs them into the classifier for the triple categorisation emotion recognition task, without on the other hand, we did not use our proposed method and input them directly into the classifier. On the other hand, we have chosen the traditional method SVM, KNN, and the deep learning supervised method MLP for the classifiers.

\begin{table}[h]
    \centering
    \begin{tabular}{@{}c@{\hskip 0.6cm}c@{\hskip 0.6cm}c@{\hskip 0.4cm}c@{\hskip 0.4cm}c@{}}
        \toprule
        \multirow{2}{*}{\hspace{0.5cm}Methods}&\multirow{2}{*}{Sampling}&\multicolumn{2}{c}{Three-Class}\\

        & & $P_{acc}$ & $P_{f}$\\
        \midrule
            \hspace{0.5cm}\multirow{2}{*}{SVM} &  w / o &  56.30& 49.53 \\
            & w & \textbf{77.69} & \textbf{76.63}\\
        \midrule
            \hspace{0.5cm}\multirow{2}{*}{KNN} &  w / o & 42.78 & 34.93\\
            & w & \textbf{62.31}& \textbf{61.00}\\
        \midrule
        \hspace{0.5cm}\multirow{2}{*}{MLP} &  w / o & 65.70 & 63.55\\
        & w  & \textbf{77.72} & \textbf{76.76}\\
        \bottomrule
        \end{tabular}

\caption{ Emotion recognition performance (\%) with the traditional classification methods using subject-independent LOOCV strategy on SEED dataset, under the conditions without (w/o) and with (w) the proposed Emotion-Agent method.}
\end{table}

For emotion recognition with the SVM classifier, the accuracy $P_{acc}$ reaches 56.30\%, and $P_f$ reaches 49.53\%. After applying our proposed method, $P_{acc}$ improves by 21.39\% and $P_f$ improves by 27.1\%. With the KNN classifier, $P_{acc}$ reaches 42.78\%, and $P_f$ reaches 34.93\%. Following the implementation of our method, $P_{acc}$ improves by 19.53\% and $P_f$ improves by 26.07\%. For the MLP classifier, $P_{acc}$ reaches 65.70\%, and $P_f$ reaches 63.55\%. After using our proposed approach, $P_{acc}$ improves by 12.02\% and $P_f$ improves by 13.21\%.
\begin{figure}[h]
    \begin{center}
        \includegraphics[width=0.45\textwidth]{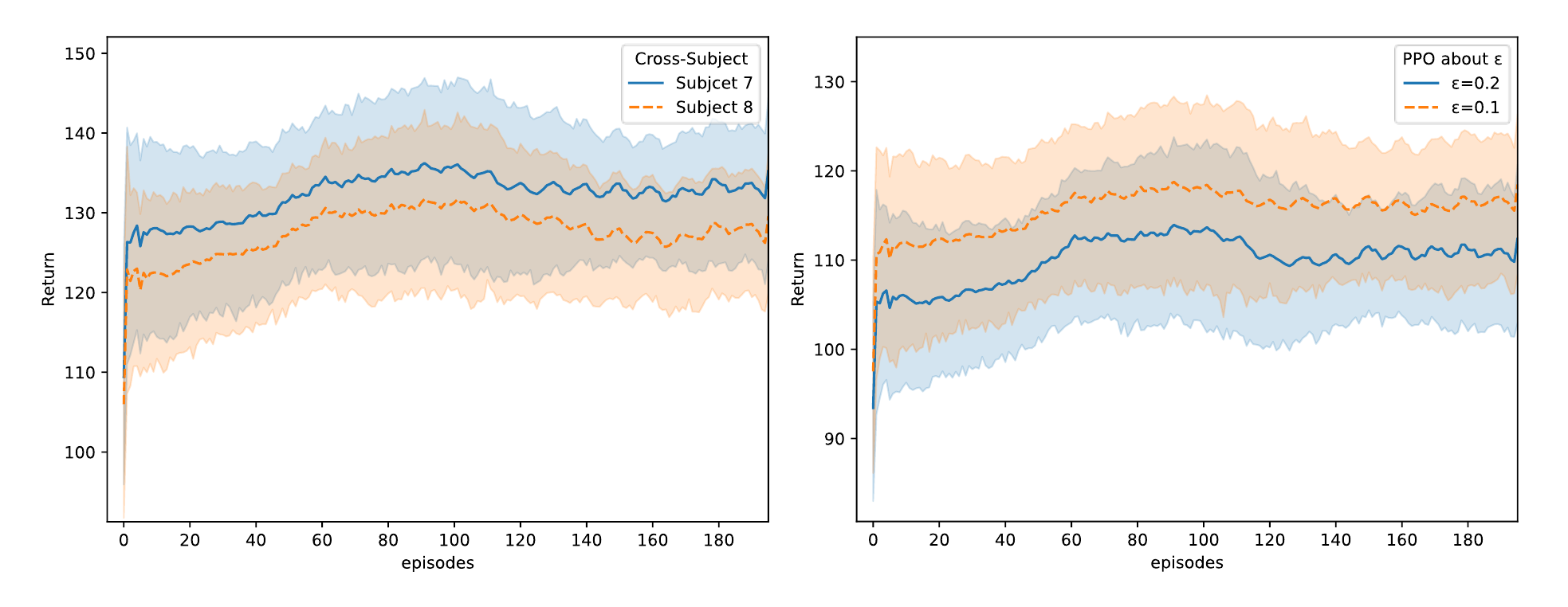}
    \end{center}
    \caption{Evaluation on Optimization Process}
    \label{fig:Percentage of Labeled Data}
\end{figure}
The experimental results show that the results on both traditional unsupervised, traditional supervised as well as deep learning supervised methods result in significant performance improvement on the SEED emotion recognition triple classification task.This experimental result also shows that reward function we designed is representative of the intensity of human emotional states to a certain extent, and this reward function allows the Agent to autonomously and unsupervised select the key segment corresponding to the intensity of the emotion. This shows, to some extent, that our model extracts emotionally rich and relevant EEG segments, and then improves the accuracy and reliability of the downstream task analysis after feeding such relevant, partially informative emotional segments into the downstream task.
}

\subsubsection{Model Optimization Process}
\label{subsec:Model Optimization Process using PPO}{
In order to have a better exploration of the EEG emotional space based on our well-designed reward function, we used PPO to complete the optimisation of the whole training process. To further study the role of the PPO algorithm in the model training process, we conducted additional experiments to explore the specific circumstances of the training process and the impact of the $\epsilon$ in the Clip operation on the model training process.

We conducte cross-subject experiments on the SEED dataset, and we compare the cumulative total return from the policy learned by the Agent during the completion of multiple Episodes of exploration and learning with the number of training sessions in the Cross-Subject experiments.Fig.2 depicts the alteration of the return during the cross-subject training of subjects 7 and 8 as the number of interaction episodes escalates. It can be observed that with a meticulously designed reward function, the cumulative benefits acquired by the agent in the task of extracting key EEG segments keep rising, and the agent progressively acquires the optimal action strategy. The right figure demonstrates the influence of the $\epsilon$  on the training process during training. It can be noted that when the upper limit of the clipping is lower, the training process is more stable and superior strategies are learned within a certain range. Constraining the ratio of new and old strategies enables the model to converge more steadily and efficiently during training.
}

\section{Conclusion}
\label{subsec:Conclusion}{
In this paper, we propose a novel unsupervised deep reinforcement learning framework, called Emotion-Agent, to automatically identify relevant and informative emotional moments from continuous EEG signals.Emotion-Agent involves unsupervised deep reinforcement learning combined with a heuristic algorithm. Constructing heuristics for reinforcement learning by constructing prior knowledge for the exploration process can dramatically improve the efficiency of intelligences in the exploration process.The extraction of fragments that have a stronger connection to emotions is more favorable for the following analysis and research.Besides according to the results, we can show the effectiveness of our proposed approach.
}

\bibliography{aaai25}

\begin{thebibliography}{39}
\providecommand{\natexlab}[1]{#1}

\bibitem[{Apostolidis et~al.(2020)Apostolidis, Adamantidou, Metsai, Mezaris, and Patras}]{apostolidis2020ac}
Apostolidis, E.; Adamantidou, E.; Metsai, A.~I.; Mezaris, V.; and Patras, I. 2020.
\newblock AC-SUM-GAN: Connecting actor-critic and generative adversarial networks for unsupervised video summarization.
\newblock \emph{IEEE Transactions on Circuits and Systems for Video Technology}, 31(8): 3278--3292.

\bibitem[{Cheng, Kolobov, and Swaminathan(2021)}]{cheng2021heuristic}
Cheng, C.-A.; Kolobov, A.; and Swaminathan, A. 2021.
\newblock Heuristic-guided reinforcement learning.
\newblock \emph{Advances in Neural Information Processing Systems}, 34: 13550--13563.

\bibitem[{Cowen and Keltner(2017)}]{cowen2017self}
Cowen, A.~S.; and Keltner, D. 2017.
\newblock Self-report captures 27 distinct categories of emotion bridged by continuous gradients.
\newblock \emph{Proceedings of the national academy of sciences}, 114(38): E7900--E7909.

\bibitem[{Duan, Zhu, and Lu(2013)}]{duan2013differential}
Duan, R.-N.; Zhu, J.-Y.; and Lu, B.-L. 2013.
\newblock Differential entropy feature for EEG-based emotion classification.
\newblock In \emph{2013 6th international IEEE/EMBS conference on neural engineering (NER)}, 81--84. IEEE.

\bibitem[{Gong et~al.(2023)Gong, Jia, Wang, Zhou, and Zhang}]{gong2023astdf}
Gong, P.; Jia, Z.; Wang, P.; Zhou, Y.; and Zhang, D. 2023.
\newblock ASTDF-Net: Attention-Based Spatial-Temporal Dual-Stream Fusion Network for EEG-Based Emotion Recognition.
\newblock In \emph{Proceedings of the 31st ACM International Conference on Multimedia}, 883--892.

\bibitem[{Hartung, Knapp, and Sinha(2011)}]{hartung2011statistical}
Hartung, J.; Knapp, G.; and Sinha, B.~K. 2011.
\newblock \emph{Statistical meta-analysis with applications}.
\newblock John Wiley \& Sons.

\bibitem[{He et~al.(2016)He, Chen, He, Gao, Li, Deng, and Ostendorf}]{he2016deepreinforcementlearningnatural}
He, J.; Chen, J.; He, X.; Gao, J.; Li, L.; Deng, L.; and Ostendorf, M. 2016.
\newblock Deep Reinforcement Learning with a Natural Language Action Space.
\newblock arXiv:1511.04636.

\bibitem[{Horikawa et~al.(2020)Horikawa, Cowen, Keltner, and Kamitani}]{horikawa2020neural}
Horikawa, T.; Cowen, A.~S.; Keltner, D.; and Kamitani, Y. 2020.
\newblock The neural representation of visually evoked emotion is high-dimensional, categorical, and distributed across transmodal brain regions.
\newblock \emph{Iscience}, 23(5).

\bibitem[{Huang et~al.(2014)Huang, Wu, Wong, and Lin}]{huang2014novel}
Huang, Y.-J.; Wu, C.-Y.; Wong, A. M.-K.; and Lin, B.-S. 2014.
\newblock Novel active comb-shaped dry electrode for EEG measurement in hairy site.
\newblock \emph{IEEE Transactions on Biomedical Engineering}, 62(1): 256--263.

\bibitem[{Ivanov and Ben-Porat(2024)}]{ivanov2024personalized}
Ivanov, D.; and Ben-Porat, O. 2024.
\newblock Personalized Reinforcement Learning with a Budget of Policies.
\newblock In \emph{Proceedings of the AAAI Conference on Artificial Intelligence}, volume~38, 12735--12743.

\bibitem[{Jiang et~al.(2023)Jiang, Zhao, Wang, and Jiang}]{jiang2023continuous}
Jiang, W.; Zhao, W.~X.; Wang, J.; and Jiang, J. 2023.
\newblock Continuous trajectory generation based on two-stage GAN.
\newblock In \emph{Proceedings of the AAAI Conference on Artificial Intelligence}, volume~37, 4374--4382.

\bibitem[{Kalashnikov et~al.(2018)Kalashnikov, Irpan, Pastor, Ibarz, Herzog, Jang, Quillen, Holly, Kalakrishnan, Vanhoucke et~al.}]{kalashnikov2018scalable}
Kalashnikov, D.; Irpan, A.; Pastor, P.; Ibarz, J.; Herzog, A.; Jang, E.; Quillen, D.; Holly, E.; Kalakrishnan, M.; Vanhoucke, V.; et~al. 2018.
\newblock Scalable deep reinforcement learning for vision-based robotic manipulation.
\newblock In \emph{Conference on robot learning}, 651--673. PMLR.

\bibitem[{Kletzander and Musliu(2023)}]{kletzander2023large}
Kletzander, L.; and Musliu, N. 2023.
\newblock Large-state reinforcement learning for hyper-heuristics.
\newblock In \emph{Proceedings of the AAAI Conference on Artificial Intelligence}, volume~37, 12444--12452.

\bibitem[{Koelstra et~al.(2011)Koelstra, Muhl, Soleymani, Lee, Yazdani, Ebrahimi, Pun, Nijholt, and Patras}]{koelstra2011deap}
Koelstra, S.; Muhl, C.; Soleymani, M.; Lee, J.-S.; Yazdani, A.; Ebrahimi, T.; Pun, T.; Nijholt, A.; and Patras, I. 2011.
\newblock Deap: A database for emotion analysis; using physiological signals.
\newblock \emph{IEEE transactions on affective computing}, 3(1): 18--31.

\bibitem[{Li et~al.(2019)Li, Qiu, Du, Wang, and He}]{li2019domain}
Li, J.; Qiu, S.; Du, C.; Wang, Y.; and He, H. 2019.
\newblock Domain adaptation for EEG emotion recognition based on latent representation similarity.
\newblock \emph{IEEE Transactions on Cognitive and Developmental Systems}, 12(2): 344--353.

\bibitem[{Li, Wang, and Lu(2021)}]{li2021multi}
Li, R.; Wang, Y.; and Lu, B.-L. 2021.
\newblock A multi-domain adaptive graph convolutional network for EEG-based emotion recognition.
\newblock In \emph{Proceedings of the 29th ACM International Conference on Multimedia}, 5565--5573.

\bibitem[{Li et~al.(2018)Li, Zheng, Cui, Zhang, and Zong}]{li2018novel}
Li, Y.; Zheng, W.; Cui, Z.; Zhang, T.; and Zong, Y. 2018.
\newblock A novel neural network model based on cerebral hemispheric asymmetry for EEG emotion recognition.
\newblock In \emph{IJCAI}, 1561--1567.

\bibitem[{Liang, Oba, and Ishii(2019)}]{liang2019unsupervised}
Liang, Z.; Oba, S.; and Ishii, S. 2019.
\newblock An unsupervised EEG decoding system for human emotion recognition.
\newblock \emph{Neural Networks}, 116: 257--268.

\bibitem[{Liang et~al.(2021)Liang, Zhou, Zhang, Li, Huang, Zhang, and Ishii}]{liang2021eegfusenet}
Liang, Z.; Zhou, R.; Zhang, L.; Li, L.; Huang, G.; Zhang, Z.; and Ishii, S. 2021.
\newblock EEGFuseNet: Hybrid unsupervised deep feature characterization and fusion for high-dimensional EEG with an application to emotion recognition.
\newblock \emph{IEEE Transactions on Neural Systems and Rehabilitation Engineering}, 29: 1913--1925.

\bibitem[{Lin, Deng, and Pan(2021)}]{lin2021combining}
Lin, Z.; Deng, Y.; and Pan, W. 2021.
\newblock Combining the strengths of inverse-variance weighting and Egger regression in Mendelian randomization using a mixture of regressions model.
\newblock \emph{PLoS genetics}, 17(11): e1009922.

\bibitem[{Liu et~al.(2024)Liu, Zhou, Wang, and Liang}]{liu2024joint}
Liu, Q.; Zhou, Z.; Wang, J.; and Liang, Z. 2024.
\newblock Joint Contrastive Learning with Feature Alignment for Cross-Corpus EEG-based Emotion Recognition.
\newblock \emph{arXiv preprint arXiv:2404.09559}.

\bibitem[{Liu et~al.(2017)Liu, Yu, Zhao, Song, Ge, and Shi}]{liu2017real}
Liu, Y.-J.; Yu, M.; Zhao, G.; Song, J.; Ge, Y.; and Shi, Y. 2017.
\newblock Real-time movie-induced discrete emotion recognition from EEG signals.
\newblock \emph{IEEE Transactions on Affective Computing}, 9(4): 550--562.

\bibitem[{Martin and Arroyo(2004)}]{martin2004agentx}
Martin, K.~N.; and Arroyo, I. 2004.
\newblock AgentX: Using reinforcement learning to improve the effectiveness of intelligent tutoring systems.
\newblock In \emph{International Conference on Intelligent Tutoring Systems}, 564--572. Springer.

\bibitem[{Ren et~al.(2023)Ren, Govil, Yang, Narasimhan, and Majumdar}]{ren2023leveraging}
Ren, A.~Z.; Govil, B.; Yang, T.-Y.; Narasimhan, K.~R.; and Majumdar, A. 2023.
\newblock Leveraging language for accelerated learning of tool manipulation.
\newblock In \emph{Conference on Robot Learning}, 1531--1541. PMLR.

\bibitem[{Schulman et~al.(2017)Schulman, Wolski, Dhariwal, Radford, and Klimov}]{schulman2017proximal}
Schulman, J.; Wolski, F.; Dhariwal, P.; Radford, A.; and Klimov, O. 2017.
\newblock Proximal policy optimization algorithms.
\newblock \emph{arXiv preprint arXiv:1707.06347}.

\bibitem[{Shi, Jiao, and Lu(2013)}]{shi2013differential}
Shi, L.-C.; Jiao, Y.-Y.; and Lu, B.-L. 2013.
\newblock Differential entropy feature for EEG-based vigilance estimation.
\newblock In \emph{2013 35th Annual International Conference of the IEEE Engineering in Medicine and Biology Society (EMBC)}, 6627--6630. IEEE.

\bibitem[{Song et~al.(2018)Song, Zheng, Song, and Cui}]{song2018eeg}
Song, T.; Zheng, W.; Song, P.; and Cui, Z. 2018.
\newblock EEG emotion recognition using dynamical graph convolutional neural networks.
\newblock \emph{IEEE Transactions on Affective Computing}, 11(3): 532--541.

\bibitem[{Vecerik et~al.(2017)Vecerik, Hester, Scholz, Wang, Pietquin, Piot, Heess, Roth{\"o}rl, Lampe, and Riedmiller}]{vecerik2017leveraging}
Vecerik, M.; Hester, T.; Scholz, J.; Wang, F.; Pietquin, O.; Piot, B.; Heess, N.; Roth{\"o}rl, T.; Lampe, T.; and Riedmiller, M. 2017.
\newblock Leveraging demonstrations for deep reinforcement learning on robotics problems with sparse rewards.
\newblock \emph{arXiv preprint arXiv:1707.08817}.

\bibitem[{Vinyals et~al.(2019)Vinyals, Babuschkin, Czarnecki, Mathieu, Dudzik, Chung, Choi, Powell, Ewalds, Georgiev et~al.}]{vinyals2019grandmaster}
Vinyals, O.; Babuschkin, I.; Czarnecki, W.~M.; Mathieu, M.; Dudzik, A.; Chung, J.; Choi, D.~H.; Powell, R.; Ewalds, T.; Georgiev, P.; et~al. 2019.
\newblock Grandmaster level in StarCraft II using multi-agent reinforcement learning.
\newblock \emph{nature}, 575(7782): 350--354.

\bibitem[{Warnell et~al.(2018)Warnell, Waytowich, Lawhern, and Stone}]{warnell2018deep}
Warnell, G.; Waytowich, N.; Lawhern, V.; and Stone, P. 2018.
\newblock Deep tamer: Interactive agent shaping in high-dimensional state spaces.
\newblock In \emph{Proceedings of the AAAI conference on artificial intelligence}, volume~32.

\bibitem[{Yarats, Kostrikov, and Fergus(2021)}]{yarats2021image}
Yarats, D.; Kostrikov, I.; and Fergus, R. 2021.
\newblock Image augmentation is all you need: Regularizing deep reinforcement learning from pixels.
\newblock In \emph{International conference on learning representations}.

\bibitem[{Zhang et~al.(2023)Zhang, Pan, Zhang, Zhang, Li, Zhang, Huang, Su, Liang, and Zhang}]{zhang2023unsupervised}
Zhang, Y.; Pan, Y.; Zhang, Y.; Zhang, M.; Li, L.; Zhang, L.; Huang, G.; Su, L.; Liang, Z.; and Zhang, Z. 2023.
\newblock Unsupervised time-aware sampling network with deep reinforcement learning for eeg-based emotion recognition.
\newblock \emph{IEEE Transactions on Affective Computing}.

\bibitem[{Zheng(2016)}]{zheng2016multichannel}
Zheng, W. 2016.
\newblock Multichannel EEG-based emotion recognition via group sparse canonical correlation analysis.
\newblock \emph{IEEE Transactions on Cognitive and Developmental Systems}, 9(3): 281--290.

\bibitem[{Zheng and Lu(2015)}]{zheng2015investigating}
Zheng, W.-L.; and Lu, B.-L. 2015.
\newblock Investigating critical frequency bands and channels for EEG-based emotion recognition with deep neural networks.
\newblock \emph{IEEE Transactions on autonomous mental development}, 7(3): 162--175.

\bibitem[{Zhong, Wang, and Miao(2020)}]{zhong2020eeg}
Zhong, P.; Wang, D.; and Miao, C. 2020.
\newblock EEG-based emotion recognition using regularized graph neural networks.
\newblock \emph{IEEE Transactions on Affective Computing}, 13(3): 1290--1301.

\bibitem[{Zhou, Qiao, and Xiang(2018)}]{zhou2018deep}
Zhou, K.; Qiao, Y.; and Xiang, T. 2018.
\newblock Deep reinforcement learning for unsupervised video summarization with diversity-representativeness reward.
\newblock In \emph{Proceedings of the AAAI conference on artificial intelligence}, volume~32.

\bibitem[{Zhou et~al.(2023)Zhou, Zhang, Fu, Zhang, Li, Huang, Li, Yang, Dong, Zhang et~al.}]{zhou2023pr}
Zhou, R.; Zhang, Z.; Fu, H.; Zhang, L.; Li, L.; Huang, G.; Li, F.; Yang, X.; Dong, Y.; Zhang, Y.-T.; et~al. 2023.
\newblock PR-PL: A novel prototypical representation based pairwise learning framework for emotion recognition using EEG signals.
\newblock \emph{IEEE Transactions on Affective Computing}, 15(2): 657--670.

\bibitem[{Zhuang et~al.(2017)Zhuang, Zeng, Tong, Zhang, Zhang, and Yan}]{zhuang2017emotion}
Zhuang, N.; Zeng, Y.; Tong, L.; Zhang, C.; Zhang, H.; and Yan, B. 2017.
\newblock Emotion recognition from EEG signals using multidimensional information in EMD domain.
\newblock \emph{BioMed research international}, 2017(1): 8317357.

\bibitem[{Zoph and Le(2017)}]{zoph2017neuralarchitecturesearchreinforcement}
Zoph, B.; and Le, Q.~V. 2017.
\newblock Neural Architecture Search with Reinforcement Learning.
\newblock arXiv:1611.01578.

\end{thebibliography}
\newpage
\section{Appendix}

\section{Datasets}
We perform related experiments on the SEED \cite{zheng2015investigating} and the DEAP \cite{koelstra2011deap}  using our proposed method Emotion-Agent. The following is a specific description of the two datasets:

\textbf{SEED}  This dataset was developed by the BCMI laboratory at Shanghai Jiao Tong University. The dataset was acquired using the 62-channel ESI NeuroScan System based on the international 10-20 system, which recorded EEG signals from subjects under different types of video stimuli.The SEED dataset acquires raw EEG signals at a sampling rate of 1000 Hz.Regarding the experimental paradigm of the SEED dataset, specifically, the EEG signals of 15 subjects (7 males and 8 females) were recorded in the SEED dataset under various video stimuli. For each subject, the video clips to be viewed were divided into three different sessions. In each session, 15 different types of film clips were involved, among which there were three types of clips that elicited different emotional states (positive, neutral, and negative moods), and each emotional state comprised five film clips.

\textbf{DEAP}  This dataset utilized a 32-channel Biosemi Active Two device with a sampling frequency of 512 Hz to record the subjects being stimulated by different one-minute-long music videos. Each video in the dataset corresponds to four labels, namely Valence, Arousal, Dominance, and Liking.A total of 32 participants in good physical condition were selected for the trial during the data collection process, consisting of 16 males and 16 females. Each subject was obligated to carry out 40 experiments, and in each of them, a 1-minute music video was watched to induce the relevant EEG. At the end of each experiment, a prompt self-assessment was conducted to rate the current state of the participant (Valence, Arousal, Dominance, and Liking), which was subsequently analysed and quantified comprehensively. Finally, a threshold value is employed to binarize the four labels for each video, thereby obtaining discrete labels for each state.

Table 1 shows the data statistics of the two datasets. We only use the preprocessed 1-s EEG signals from session 1 for both datasets.
When conducting experiments with our model, we use the preprocessed 1-s EEG signals of session 1 for the SEED dataset. For the DEAP dataset, we likewise employ the preprocessed 1-s EEG signals.

\section{Preprocessing}
In the preprocessing section, we will respectively introduce the pre-processing of the two datasets and the extraction of the Differential Entropy Feature (DE feature) \cite{duan2013differential} corresponding to the EEG signals.

The raw SEED dataset was initially preprocessed. To be specific, the raw EEG data were initially downsampled to a sampling rate of 200 Hz and filtered through a 1-75 Hz bandpass filter to filter out noise and eliminate artefacts. Next, the preprocessed EEG signals were divided into multiple segments by utilizing a sliding window with a length of 15 to obtain the EEG signals after the preliminary processing.
In an effort to obtain features in the EEG signals that are more closely related to the brain state, differential entropy features were extracted for the EEG signals measured in seconds (with a 200 Hz sampling rate corresponding to 200 sampling points) using a band-pass filter ($\delta$ wave 0.5-4 Hz, $\theta$ wave: 4-8 Hz, $\alpha$ wave: 8-13 Hz , $\theta$ wave: 13-32 Hz, $\gamma$ wave: 32-50 Hz). The specific expression for calculating the differential entropy of EEG signals is as follows:

\begin{equation}
h(X)=\frac{1}{2}\log(2\pi e\sigma^{2}),
\end{equation}
where the time series $X$ obeys the Gauss distribution $N(\mu,\sigma^2)$. It has been proven that, for a fixed length EEG sequence, DE is equivalent to the logarithm ES in a certain frequency band \cite{shi2013differential}. DE was employed to construct features in five frequency bands mentioned above.

\begin{table*}[h]
    \centering
    \begin{tabular}{@{}c@{\hskip 0.2cm}c@{\hskip 0.2cm}c@{\hskip 0.2cm}c@{\hskip 0.2cm}c@{\hskip 0.2cm}c@{\hskip 0.2cm}c@{\hskip 0.3cm}c@{\hskip 0.3cm}c@{}}
        \toprule
        \midrule
        & Datasets & Subject & Sessions & Trials & Channels & Sampling Rate (Hz) & \multicolumn{2}{c}{Classes}\\
        \midrule
        & SEED & 15 & 3 & 15 & 62 & 1000 & \multicolumn{2}{c}{3(Negative, Neutral, Positive)}\\ 
        & DEAP & 32 & 1 & 40 & 32 & 512 & 2 Valence (Negative, Positive) & 2 Arousal (Calm, Active) \\
        \midrule
        \bottomrule
        \end{tabular}
\label{tab:Classification accuracies for different masking strategies.}
\caption{Detailed description of the SEED and DEAP datasets.}
\end{table*}

After processing the EEG signal per second through differential entropy, the feature dimension changes from (62, 200) to (62, 5), where 62 represents the number of device channels. From the sample features per second, we extracted the features in 5 frequency bands of each channel and flattened them. Thus, the feature dimension of the sample features per second becomes (1, 310). We use the DE features of the EEG signal per second as the input of the model, which corresponds to three emotion labels (negative, neutral, and positive).

For the DEAP dataset, the original sampling frequency was 512 Hz. Subsequently, the data was downsampled to 128 Hz, while removing artefacts and deleting the first three seconds of silence in each experiment to obtain the initially processed EEG signal. Similarly for this dataset, to extract DE features ($\delta$ wave 0.5-4 Hz, $\theta$ wave: 4-8 Hz, $\alpha$ wave: 8-14 Hz , $\theta$ wave: 14-32 Hz, $\gamma$ wave: 32-50 Hz), the feature dimension per second changes from (32, 128) to (32, 5). We perform a flatten operation on it and the sample feature dimension per second is altered to (1, 160) to obtain the DE features of the EEG. We use the proposed model in Valence and Arousal two labels for experimentation, and both labels correspond to binary classification tasks.

\section{Implementation Details}
In this section, We will provide a detailed account of the specificities of the two processes, (1) Prototype Learning, and (2) Reinforcement Learning, that are employed in the model when training the model for cross-subject experiments (Subject-Leave-One-Out Cross-Validation). In order to better describe the training details, We define the total number of subjects in the dataset as $N$.

\subsection{Prototype Learning}
In the prototype learning stage, with the aim of obtaining a global overview of the data distribution, we clustere the data of $N-1$ subjects through the utilization of the heuristic algorithm K-Means. In this case, the hyperparameter $n\_clusters$ of K-Means was set to the total number of emotion categories (set to 3 for the SEED dataset and 2 for the DEAP dataset). After several iterations of the algorithm, we obtain the emotion prototype. Additionally, we compute the mean of the sum of squared errors in each cluster as the variance of the data distribution within each cluster. We pass the sentiment prototypes $\mu_c$ for each emotion category and the variance $\sigma^2$ within each cluster to the second stage of learning.

Additionally, we employ the labels generated during the unsupervised clustering process as semantic information for subsequent EEG features, and thereby we define such a space as \textbf{Emotional Space}.

\subsection{Reinforcement Learning}
In the reinforcement learning stag, we require Trial-Level EEG data for delineation. The training data consist of the EEG DE features of a subject conducting an experiment to complete an indefinitely long sequence of actions in this manner as a decision-making process, where the action space is the discrete action $A = \{0, 1\}$.

The actor's network comprises one layer of LSTM where the number of hidden layer nodes is 128 and two fully connected layers where the hidden nodes range from $256\rightarrow128, 128\rightarrow2$. The network setup of the Critic is one layer of LSTM with 128 nodes and two fully connected layers where the hidden layer nodes in the fully connected layers are $256\rightarrow128$, $128\rightarrow1$. The actor is optimized by Adam's optimizer and the learning rate is set to 1e-4, while the Critic's learning rate is also optimized by Adam's optimizer and is set to 1e-3. Additionally, in the PPO algorithm, $\gamma$ is set to 0.98, $\lambda$ is set to 0.95, and $\epsilon$ is set to 0.2 for the algorithms (refer to the original Eq. (\ref{equation: ratio}), Eq. (\ref{equation: GAE}), Eq. (\ref{equation: object})). All experiments are carried out using PyTorch 1.13.1 on an NVIDIA GeForce RTX 3090 GPU. 

\subsection{Distribution-Prototype Reward}
We obtain information about the global distribution based on the prototype learning stage. We use the mean and variance of each emotion cluster to reflect the individual variability within the distribution of EEG features for each emotion category. We propose two reward functions based on the distributional information derived from prototype learning: center reward and inter-intra reward.

We incorporate the relevant theory of \textbf{Inverse Variance Weighting}  \cite{hartung2011statistical} presented in Inter-Intra reward here. \\
If a series of independent measurements of a random variable are represented by $y_i$ and possess a variance of $\sigma_i^2$, then the inverse variance weighted average of these measurements is:

\begin{equation}
    \hat{y}=\frac{\sum_iy_i/\sigma_i^2}{\sum_i1/\sigma_i^2}.
\end{equation}
Among all the methods of weighted averaging, the inverse variance weighted average has the least variance. The expression of its variance is as follows:
\begin{equation}
    D^2(\hat{y})=\frac{1}{\sum_i1/\sigma_i^2}.
\end{equation}
If the variances of the measurements are equalized, the inverse variance weighted average if the same as the simple average.

\section{Additional Results}
To further explore the EEG key segments extracted by our proposed method Emotion-Agent, we conduct additional experiments on the SEED dataset using the proposed model, visualized the extracted EEG keyfragments using t-SNE, and compared them with the case without.

\begin{figure}[h]
    \begin{center}
        \includegraphics[width=0.45\textwidth]{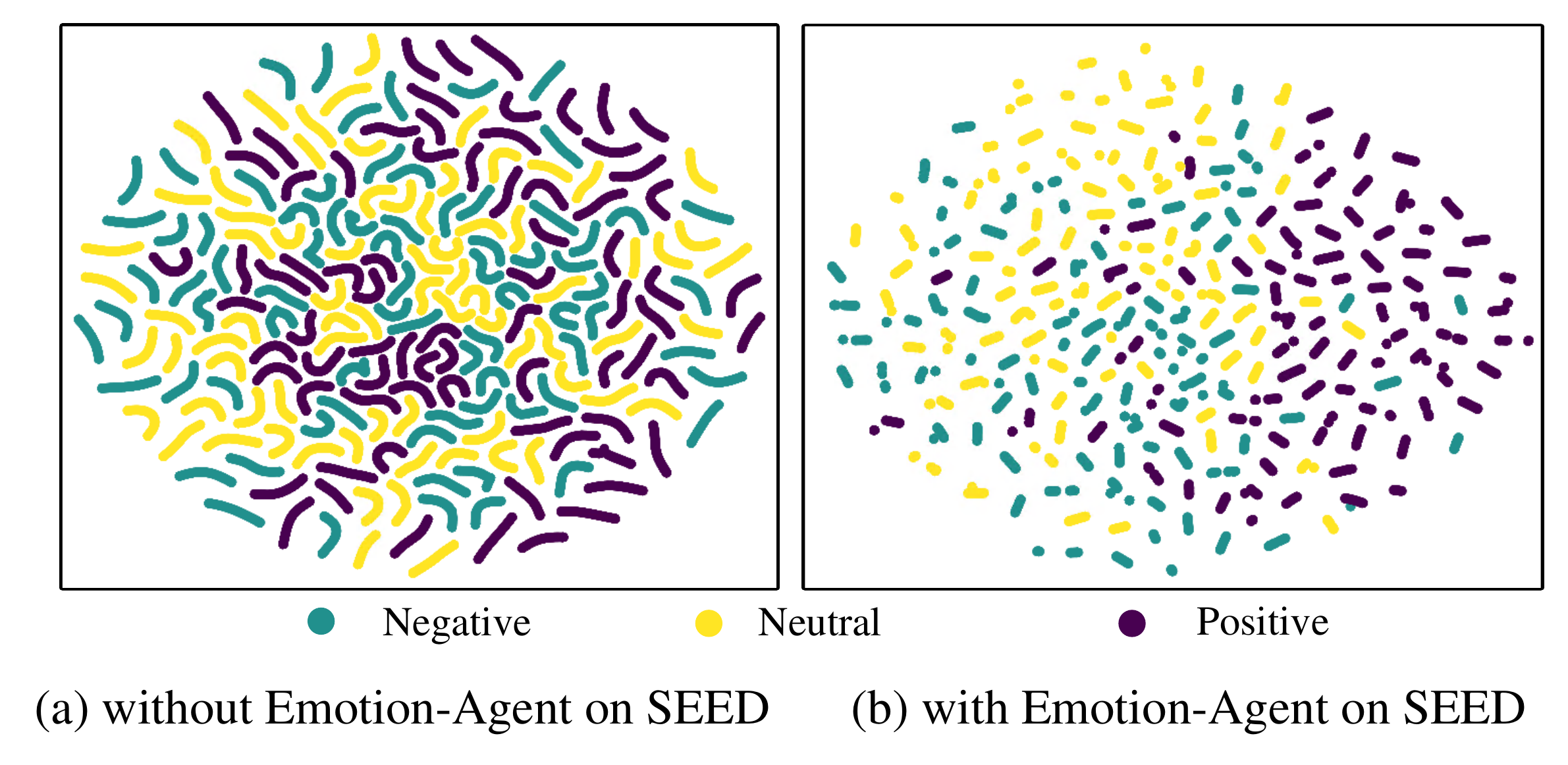}
    \end{center}
    \caption{Comparsion of data distribution using t-SNE with and without Emotion-Agent on the SEED dataset.}
    \label{fig:TSNE}
\end{figure}

\begin{figure}[h]
    \begin{center}
        \includegraphics[width=0.45\textwidth]{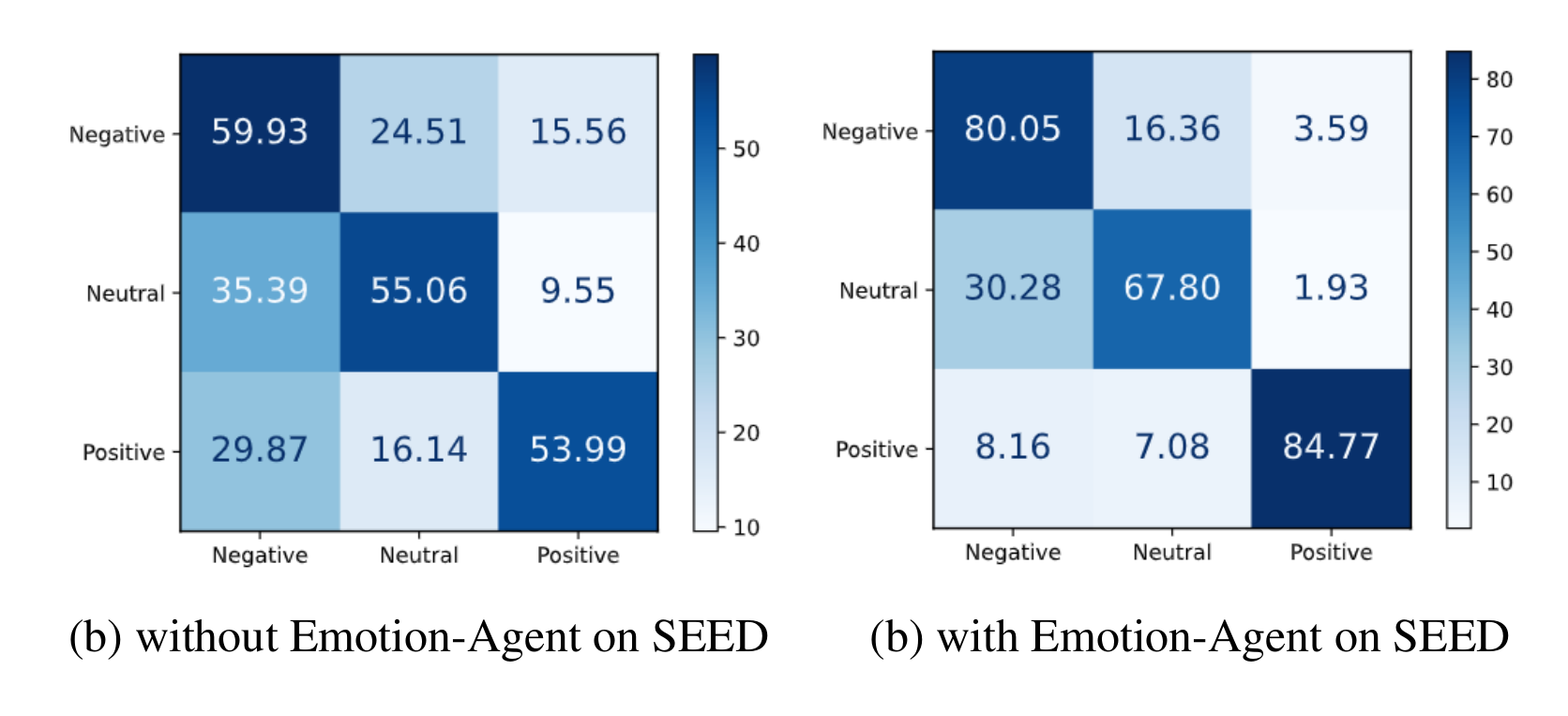}
    \end{center}
    \caption{
    Comparsion of confusion matrices for emotion recognition with and without Emotion-Agent on the SEED dataset.}
    \label{fig:Percentage of Labeled Data}
\end{figure}

Figure 1 compares the with and without Emotion-Agent models, with SEED DE features as input, and visually compares the original DE features with the data after extracting key segments through t-SNE results. It can be seen that after extracting key segments, the separability of the entire data distribution is improved compared to before, and the same conclusion can be drawn from the improvement in classification accuracy.

Figure 2 compares the with and without Emotion-Agent models, with the SEED DE features as input and the classification confusion matrix obtained using SVM as the classifier. The experimental results show that compared with without, with achieves higher classification accuracy on all three emotion categories.\\

\end{document}